\documentclass[11pt,a4paper, linktocpage,colorlinks=true, linkcolor=bluscuro,urlcolor=bluscuro,citecolor=bluscuro]{article}

\usepackage{jheppub}
\usepackage{graphicx}
\usepackage{bm}
\usepackage{xcolor}
\usepackage{booktabs}
\PassOptionsToPackage{breaklinks}{hyperref}
\definecolor{bluscuro}{rgb}{0.15, 0.2, .85}
\usepackage{amsmath, amssymb}
\usepackage{slashed}
\usepackage{latexsym}
\usepackage{braket}
\usepackage[export]{adjustbox}
\usepackage{mathrsfs}
\usepackage{color}
\usepackage{multirow}
\usepackage{xspace}
\usepackage{comment}
\usepackage[caption=false]{subfig}
\usepackage{cancel}
\usepackage{url}
\usepackage{nicematrix}
\usepackage{physics}
\usepackage{pdfpages}
\usepackage{float}

\newcommand{\ddd}{\mathrm{d}}

\title{The gyromagnetic factor of charged rotating black holes in various dimensions from scattering amplitudes}

\author[a, b]{Claudio Gambino,}
\author[b]{Fabio Riccioni,}
\author[a]{Victor Sanz Sanchis}

\affiliation[a]{Dipartimento di Fisica,  Universit\`a di Roma ``La Sapienza"}
\affiliation[b]{Sezione INFN Roma1, Piazzale Aldo Moro, 00184, Roma, Italy}
\emailAdd{claudio.gambino@uniroma1.it. Fabio.Riccioni@roma1.infn.it, sanzsanchis.2043916@studenti.uniroma1.it}
\abstract{Classical black hole spacetimes can be recovered from the classical limit of quantum scattering amplitudes in a low-energy effective field theory of gravity.
In this work we compute, at first post-Minkowskian and dipole order,  the metric and the electromagnetic potential for charged and rotating black holes in general spacetime dimensions from amplitudes describing the emission of either a graviton or a photon from a massive and charged Dirac fermion field up to one loop.
In addition, we introduce a Pauli non-minimal coupling, to parametrize the black hole's gyromagnetic factor $\mathfrak{g}$.
We are able to reproduce the Kerr-Newman solution in four dimensions, as well as the Chong-Cvetič-Lü-Pope solution, from five-dimensional supergravity, which includes a Chern-Simons interaction.
Crucially, we show that for a charged Myers-Perry like black hole in $d+1$ spacetime dimensions, its gyromagnetic factor is equal to $\mathfrak{g}=(d-1)/(d-2)$.
Hence, only in $3+1$ dimensions minimal coupling is sufficient to describe black holes from scattering amplitudes.}

\begin{document}
\maketitle

\section{Introduction}
One of the most exciting predictions of Einstein's theory of gravitation is the existence of black holes (BHs).
Today, thanks to both strong observational evidence and extensive theoretical developments over the past century, we have a detailed understanding of their properties and the various physical processes they are involved in.

In $3+1$ dimensions, BHs within the Einstein-Maxwell theory exhibit a number of remarkable features.
A series of uniqueness theorems shows that any isolated, stationary, regular BH with spherical horizon topology belongs to the Kerr-Newman family~\cite{Bardeen:1973gs,Robinson:1975bv,Robinson:2004zz,Cardoso:2016ryw}.
These solutions are uniquely determined by their global charges measured at infinity, which also serve to define their multipole moments~\cite{Geroch:1970cd,Thorne:1980ru,Hansen:1974zz}.
These multipoles are then closely related to the post-Minkowskian (PM) expansion, namely an expansion in powers of Newton's gravitational constant $G$~\cite{Bern:2019nnu,Damour:2016gwp,Gambino:2024uge}.

The quantities that characterize these solutions are the mass, electric charge and angular momentum~\cite{Misner:1973prb,Kleihaus:2007kc,Cardoso:2016ryw}.
While all BHs possess mass, switching on or off the electric charge and angular momentum yields the four standard BH solutions in $3+1$ dimensions.
The simplest case, with no charge and no rotation, is the Schwarzschild metric~\cite{Schwarzschild:1916uq}.
Introducing an electric charge yields the Reissner–Nordstr\"om solution~\cite{Reissner:1916cle}, while including angular momentum leads to the rotating Kerr solution~\cite{Kerr:1963ud}.
Finally, combining both charge and rotation gives the Kerr–Newman spacetime~\cite{Newman:1965my}.

In contrast to $3+1$ dimensions, in higher-dimensional spacetime the BH uniqueness theorems no longer hold, and solutions with different horizon topologies can be found~\cite{Emparan:2008eg,Cardoso:2016ryw}.
For instance, in $4+1$ dimensions, in a specific parameter space region of mass and angular momentum, one can construct Myers-Perry BHs with an $S^3$ horizon, and black rings with a $S^2\times S^1$ horizon~\cite{Myers:1986un,Emparan:2001wn}. 

The extension to higher dimensions of asymptotically-flat BHs with spherical horizons began with the seminal work of Tangherlini~\cite{Tangherlini:1963bw}, who generalized the metric for a static electrically charged BH, which we refer to as the Reissner-Nordstr\"om-Tangherlini solution, in arbitrary dimensions. 
For non-static BHs, the key breakthrough came from Myers and Perry~\cite{Myers:1986un}, who constructed solutions for uncharged, rotating BHs by decomposing their rotation into independent planes.
However, exact charged rotating solutions in pure higher-dimensional Einstein-Maxwell theory are not known in closed form~\cite{Kunz:2017pnm}, and in~\cite{Ortaggio:2023rzp} it is shown that such solutions cannot exist within the Kerr-Schild gauge. 
Perturbative constructions have been developed in $4+1$~\cite{Aliev:2004ec,Navarro-Lerida:2010orf} and arbitrary dimensions~\cite{Aliev:2006yk}, and in theories with additional fields or interactions one can find exact and perturbative approaches to construct these families of charged solutions~\cite{Horowitz:1995tm,Llatas:1996gh,Youm:1997hw,Chong:2005hr,Kunz:2006jd,Kleihaus:2007kc,Blazquez-Salcedo:2013wka,Kunz:2017pnm,Ortaggio:2023rzp,Deshpande:2024vbn}. In particular, in $4+1$ dimensions, the inclusion of a specif Chern-Simons term allows to construct an exact BH metric known as the Chong-Cvetič-Lü-Pope (CCLP) solution~\cite{Chong:2005hr}. 
For some of the most recent analyses on this topic see Refs. \cite{Ortaggio:2023rzp,Deshpande:2024vbn}.

In this work we reconstruct the far-field behavior of both the metric and the electromagnetic potential for charged, rotating sources by matching the classical limit $(\hbar \to 0)$ of scattering amplitudes in an effective field theory (EFT) of gravity to the PM expansion of classical BH solutions. In four dimensions, it is known that the Kerr-Newman metric results from scattering amplitudes of charged particles minimally coupled to gravity \cite{Chung:2019yfs,Moynihan:2019bor}. This extends to charged objects the result of \cite{Chung:2018kqs}, showing that
 the energy-momentum tensor of a massive higher spin field minimally coupled to gravity, in the limit in which the spin goes to infinity, exactly reproduces in the classical limit all the multipoles of a Kerr black hole. In higher dimensions, the Tangherlini metric was derived from the graviton emission of massive scalars \cite{Mougiakakos:2020laz}, and the result was then extended to the Reissner-Nordstr\"om-Tangherlini solution in \cite{DOnofrio:2022cvn} by considering charged scalars. As far as rotating objects are concerned, in \cite{Gambino:2024uge} it was shown that the Myers-Perry solution up to quadrupole order results from the scattering of spin-1 fields with gravity, where an additional non-minimal coupling has to be included in order to reproduce the correct quadrupole term.
It is important to point out that this analysis is clearly perturbative in nature, in the sense that the PM expansion corresponds to an expansion in graviton loops, so that the amplitude without such loops gives the metric and electromagnetic potential at first PM order.  Moreover, by considering a massive field with a given spin, one simply truncates the multipole expansion at the corresponding given order. For instance, in four dimensions, by considering a minimally coupled spin 1 field one reproduces the Kerr solution up to quadrupole order, while an additional higher derivative term can be added to reproduce the quadrupole of any rotating object with spin-induced multipoles \cite{Gambino:2024uge}.

Our aim here is to extend the analysis above to charged rotating solutions in arbitrary dimensions. Since we are interested in the dipole terms of the solution, 
we model the source as a massive, electrically charged spin-$\tfrac{1}{2}$ field, which naturally encodes both mass and angular momentum through its energy-momentum tensor and spin structure.
Beyond minimal coupling, we incorporate a Pauli-type term, that allows us to track how additional operators affect the gravitational and electromagnetic multipolar structure of the solution. This is the only non-minimal term that can contribute at dipole order. 
This amplitude-based setup enables us to define a higher-dimensional gyromagnetic factor for BHs and directly compare it with known metrics like Kerr–Newman or CCLP solutions. The final outcome is that any possible electrically charged extension of the Myers-Perry solution in $d+1$ dimensions gives a gyromagnetic factor $\mathfrak{g}=(d-1)/(d-2)$. From the amplitude point of view, this implies that in higher dimensions one must include a Pauli term and only in $3+1$ dimensions a minimal coupling is sufficient to describe a charged black hole. 

\textbf{Conventions.}
We work in the mostly negative signature with $\eta_{0 0} = +1$ and in natural units, $\hbar=c=\varepsilon_0=1$, whereas we keep the gravitational coupling constant $G$ explicit.
Greek indices are for spacetime components and Latin indices are for space components only, and $d+1$ is the number
of spacetime dimensions.

\section{Metric and electromagnetic potential from scattering amplitudes}

We start from the action of a massive Dirac field $\psi$ of mass $m$ and charge $Q$, minimally coupled to gravity and electromagnetism in generic $d+1$ dimensions
\begin{equation}\label{eq:minimal_general_action}
    \mathcal{S}
    = \int \ddd^{d+1} x ~\mathrm{e}\,\mathcal{L}
    = \int \ddd^{d+1} x ~\mathrm{e} \left( -\frac{2}{\kappa^2}R-\frac{1}{4}F_{\mu \nu} F^{\mu \nu}  +\bar{\psi}\left(i\,e^{\mu}_{~\alpha}\gamma^{\alpha} D_\mu-m \right)\psi\right)\ ,
\end{equation}
with $\kappa=\sqrt{32 \pi G}$, $g_{\mu \nu}=\eta_{\alpha \beta}\,e^{\alpha}_{~\mu}\,e^{\beta}_{~\nu}$, $\mathrm{e}=|\det e^{\alpha}_{~\mu}|=\sqrt{|\det g_{\mu \nu}|}$, and the covariant derivative, defined as\footnote{Note that the spin-connection term does not contribute in the evaluation of the three-point vertex.}
\begin{equation}\label{eq:covariant_der_def}
    D_\mu\psi=\partial_\mu \psi+ i Q A_\mu \psi - \frac{1}{2} \omega_{\mu \alpha \beta}  \Sigma^{\alpha \beta} \psi\ ,
\end{equation}
where $\Sigma^{\alpha \beta}=\frac{i}{4} [\gamma^\alpha,\gamma^\beta]$ are the generators of the Lorentz group in the spin-$\frac{1}{2}$ representation.

Then, we consider the PM expansion of the metric $g_{\mu \nu}$ and the electromagnetic potential $A_{\mu}$, around flat space as
\begin{equation}\label{eq:pm_exp_a_h}
    g_{\mu \nu}(x) = \eta_{\mu \nu} + \kappa\,h_{\mu \nu}(x) = \eta_{\mu \nu}+\kappa \sum_{n\geq1}h^{(n)}_{\mu \nu}(x)\ , \qquad A_\mu(x) = \sum_{n \geq 0} A^{(n)}_{\mu}(x)\ ,
\end{equation}
where $h_{\mu \nu}^{(n)}(x)$ and $A^{(n)}_{\mu}(x)$ are the terms corresponding to the $n$-th power of $G$.

In order to fix the gauge, we adopt the Harmonic (de Donder) gauge for the metric and the Feynman gauge for the electromagnetic potential.
Therefore, at each $n$PM order, the linearized Einstein's equations, in momentum space and assuming a stationary source, reduce to~\cite{Donoghue:1994dn,Mougiakakos:2020laz}
\begin{equation}\label{eq:eins_eq_exp_pm_momentum}
     \kappa\,h_{\mu \nu}^{(n)}(\vec{x}) = -16\pi G \int \frac{\ddd ^d \vec{q}}{(2\pi)^d}\frac{e^{i \vec{q}\cdot\vec{x}}}{\vec{q}^{\,2}}\left( T_{\mu \nu}^{(n-1)}(\vec{q}\,)-\frac{1}{d-1}\eta_{\mu \nu}\, T^{(n-1)}(\vec{q}\,) \right)\ ,
\end{equation}
for the metric, and~\cite{DOnofrio:2022cvn}
\begin{equation}\label{eq:em_equations_pm_momentum}
    A_{\mu}^{(n)}(\vec{x}) = \int \frac{\ddd ^d \vec{q}}{(2\pi)^d} \frac{e^{i \vec{q}\cdot\vec{x}}}{\vec{q}^{\,2}}\, j^{(n)}_{\mu}(\vec{q}\,)\ ,
\end{equation}
for the gauge potential.
Here, $T_{\mu \nu}^{(n)}$ is the energy-momentum tensor (EMT) at $n$PM order, with $T=\eta^{\mu \nu}T_{\mu \nu}$, and $j^{(n)}_{\mu}$ is the electromagnetic current.

The classical EMT and electromagnetic current are obtained from taking the classical limit of $3$-point graviton and photon emission amplitudes.
Following the idea outlined in~\cite{Bjerrum-Bohr:2002fji, Mougiakakos:2020laz, DOnofrio:2022cvn,Gambino:2024uge}, diagrams at $l$-loop order with $n_g$ graviton and $n_p$ photon inserted into the massive line satisfy $l = n_g + n_p -1$.
Therefore, the classical contribution to the EMT can be computed as a loop-wise expansion involving diagrams of the following form~\cite{Gambino:2024uge}
\begin{equation}\label{eq:metric_class_diagram_general}
\includegraphics[valign=c,width=0.45\textwidth]{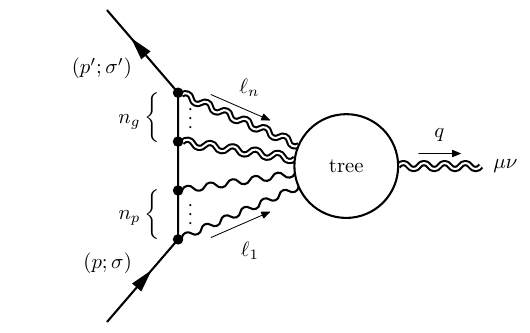}
\,=-i\frac{\kappa}{2}~ T^{(n_g+\frac{n_p}{2}-1)}_{\mu \nu}(\vec{q}\,)~\delta_{\sigma\sigma'}\ ,
\end{equation}
whereas the electromagnetic current follows from~\cite{DOnofrio:2022cvn}
\begin{equation}\label{eq:potential_class_diagram_general}
\includegraphics[valign=c,width=0.45\textwidth]{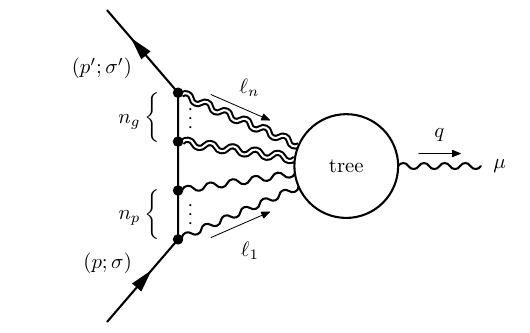}
\,=-i~ j^{(n_g+\frac{n_p+1}{2}-1)}_{\mu}(\vec{q}\,)~\delta_{\sigma\sigma'}\ ,
\end{equation}
where $\sigma$ and $\sigma'$ are the physical polarizations of the particles and $q=p'-p$ is the transferred momentum.
Indeed, one can see that the classical limit of the EMT, and the current, reduce to terms proportional to the $l$-loop master integral~\cite{Donoghue:1994dn,Holstein:2004dn,Mougiakakos:2020laz}
\begin{equation}\label{eq:master_integral_def}
    J_{(l)}(\vec{q}\,) = \int \prod_{i=1}^{l} \frac{\ddd^d \vec{\ell}_i}{(2 \pi)^d}\frac{\vec{q}^{\,2}}{\left( \prod_{i = 1}^{l} \vec{\ell}_i^{\,2} \right)\,\left(\vec{q}-\sum_{i=1}^{l} \vec{\ell}_i  \right)^2}\ ,
\end{equation}
whose Fourier transform yields the far-field expressions for the metric and potential.

From General Relativity it is known that in the gravitational multipole expansion, the quadrupole term is the first that discriminates between different solutions of the Einstein equations, so up to dipole order, the multipole expansion is unique~\cite{Thorne:1980ru}.
From an amplitude point of view, this implies that up to dipole order, classical contributions are not affected by the inclusion of additional non-minimal couplings.
However, for electromagnetic multipoles this is not true. 
Indeed, the electromagnetic dipole term is not unique, and one can see that there exists a single non-minimal Pauli coupling~\cite{Pauli:1941zz}
\begin{equation}\label{eq:non_minimal_pauli_lagrangian}
{}^{\mathrm{nm}}\mathcal{L}_{\mathrm{\psi^2 A}}
    = -i\,\zeta\frac{Q}{2 m}\,\bar{\psi}\Sigma^{\mu \nu}\psi\,F_{\mu \nu}\ ,
\end{equation}
where $\zeta$ is a dimensionless parameter, that modifies the fermion-photon vertex that contributes in the classical limit at this order.

\begin{figure}[h]
\centering
\begin{equation*}
\includegraphics[valign=c,width=0.25\textwidth]{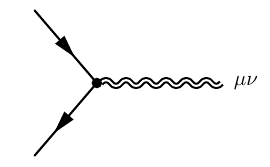}+\includegraphics[valign=c,width=0.25\textwidth]{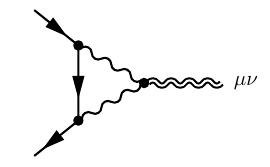}+2\times\includegraphics[valign=c,width=0.25\textwidth]{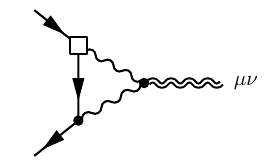}
\end{equation*}
\caption{Diagrams, with corresponding multiplicity, employed for the computation of the PM expansion of the metric up to 1PM and in dipole approximation. The fermion-fermion squared vertex represents the Pauli coupling.}
    \label{fig:metric_one}
\end{figure}

\begin{figure}[h]
    \centering
\begin{equation*}
\includegraphics[valign=c,width=0.2\textwidth]{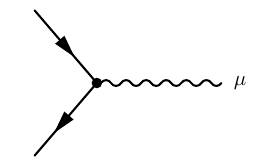}+\includegraphics[valign=c,width=0.2\textwidth]{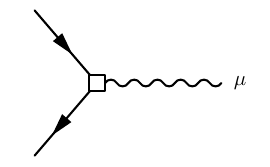}+2\times\includegraphics[valign=c,width=0.2\textwidth]{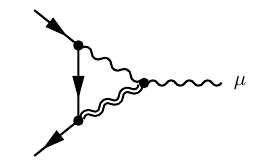}+2\times\includegraphics[valign=c,width=0.2\textwidth]{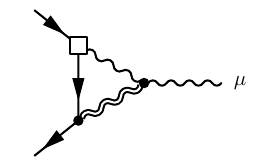}
\end{equation*}
\caption{Diagrams, with corresponding multiplicity, used in the calculation of the PM expansion of the potential up to 1PM, linear in the charge and in dipole approximation. The fermion-fermion squared vertex represents the Pauli coupling.}
    \label{fig:potential_one}
\end{figure}

In this work we derive the PM expansion up to first order for both the metric and the gauge potential.
Specifically, this corresponds to evaluating only the graviton-emission diagrams  in Fig.~\ref{fig:metric_one} and the photon-emission diagrams in Fig.~\ref{fig:potential_one}  that do not contain graviton loops.\footnote{Note that this also implies that our analysis is not affected if one modifies the graviton interactions, like for instance in Gauss-Bonnet theories.} From such diagrams we compute the metric and  the electromagnetic potential respectively.

The details on extracting the tree-level contribution of the metric in the chargeless spin-$\frac{1}{2}$ case can be found in~\cite{Bjerrum-Bohr:2002fji}, as well as the general procedure to compute the various loop diagrams. Recalling the definition of the dressed vertex 
\begin{equation}
    \bra{p',\sigma'}\tau_{\mu}\ket{p,\sigma} =\hat{\tau}_{\mu}\delta_{\sigma\sigma'}+\mathcal{O}(\hbar)\ ,
\end{equation}
the vertex associated to the minimal interaction term is
\begin{equation}\label{eq:01_current}
    \includegraphics[valign=c,width=0.2\textwidth]{Diagrams_PDF/Photon_Minimal_Vertex.pdf}=\quad ( \hat{\tau}_{\psi^2 A})_{\mu} = -i\left(Q\,\delta_\mu^0-i\,Q\,S_{\mu \nu}\,q^\nu\right)\ ,
\end{equation}
while the one coming from the Pauli coupling is \eqref{eq:non_minimal_pauli_lagrangian}
\begin{equation}\label{eq:01_current_nm}
    \includegraphics[valign=c,width=0.2\textwidth]{Diagrams_PDF/Photon_Pauli_Vertex.pdf}=\Big({}^{\mathrm{nm}}\hat{\tau}_{\psi^2 A}\Big)_{\mu} = - \zeta\,Q\,S_{\mu \nu}\,q^\nu\ ,
\end{equation}
where to derive Eqs.~\eqref{eq:01_current} and \eqref{eq:01_current_nm} we employed the Gordon identity~\cite{Itzykson:1980rh}
\begin{equation}\label{eq:gordon_identity}
    \bra{p',\sigma'} \gamma_\mu \ket{p,\sigma} = \bra{p',\sigma'} \left( \frac{p_\mu+p'_\mu}{2m} - i\frac{p^\nu-p'^\nu}{m} \Sigma_{\mu \nu} \right)\ket{p,\sigma}\ ,
\end{equation}
and the relations~\cite{Kosower:2018adc,Bern:2020buy,Chung:2018kqs}
\begin{equation}\label{eq:braket_states_classical}
        \braket{p',\sigma'}{p,\sigma} = \delta_{\sigma \sigma'} + \mathcal{O}\left(\hbar\right)\ ,
         \qquad
         \bra{p',\sigma'}\Sigma_{\mu \nu}\ket{p,\sigma} = m\, S_{\mu \nu}\, \delta_{\sigma \sigma'} + \mathcal{O}\left(\hbar^0\right)\ ,
\end{equation}
with $S_{\mu \nu}$ being the anti-symmetric spin-density tensor of the classical source.
From expressions \eqref{eq:01_current} and \eqref{eq:01_current_nm}, the leading order of the gauge potential is
\begin{equation}\label{eq:potential_with_zeta}
    \begin{aligned}
        A^{(0)}_{0}(\vec{x}) &= \frac{Q}{4\,\pi}\,\rho(r)\ ,
        \\
        A^{(0)}_{i}(\vec{x}) &= (1+\zeta)(d-2)\frac{Q}{4\,\pi \,r^2}\,\rho(r)S_{ik}\,x^k\ ,
    \end{aligned}
\end{equation}
where $\rho(r)=\frac{4\,\pi\,r^{2-d}}{(d-2)\,\Omega_{d-1}}$, with $r^2=x_1^2+\dots+x_d^2$ and $\Omega_{d-1} = \frac{2 \pi^{d/2}}{\Gamma(d/2)}$ being the area of the unit $(d-1)$-sphere.

Then, rewriting Eq.~\eqref{eq:potential_with_zeta} explicitly
\begin{equation}\label{eq:magnetic_zeta_d}
    A^{(0)}_{i}(\vec{x}) = (1+\zeta) \,\frac{1}{\Omega_{d-1}~r^d} Q S_{ik}\,x^k\ ,
\end{equation}
allows us to compare it with the generic form of an electromagnetic potential at dipole order, given by
\begin{equation}\label{eq:dipole_potential}
    A^{\mathrm{dip.}}_i = \frac{\mathfrak{g}}{2}\frac{1}{\Omega_{d-1}\,r^d}\,Q S_{i k}\,x^k\ ,
\end{equation}
where $\mathfrak{g}$ is the gyromagnetic factor in $d+1$ dimensions.
By direct comparison we see that the total gyromagnetic factor of a spin-$\frac{1}{2}$ fermion in a Dirac-Pauli EFT is
\begin{equation}
    \mathfrak{g}_{\mathrm{Dirac-Pauli}}=2\left(1+
\zeta\right)\ .
\end{equation}
From now on we will write our results keeping the factor $\mathfrak{g}$ explicit.

The next step is to compute the loop amplitudes with multiple graviton and photon insertions on the massive line shown in both Fig.~\ref{fig:metric_one} and Fig.~\ref{fig:potential_one}, following the procedure covered in detail in~\cite{Donoghue:2001qc,Holstein:2004dn,Bjerrum-Bohr:2018xdl,Mougiakakos:2020laz,DOnofrio:2022cvn,Gambino:2024uge}.
All together, the resulting far-field metric in general $d+1$ dimensions up to 1PM and dipole order, within Einstein-Maxwell theory with a Pauli coupling, takes the form
\begin{equation}\label{eq:metric_d_dimensions}
    \begin{aligned}
        \kappa\,h_{0 0}(\vec{x}) &= -4\frac{d-2}{d-1}G m\,\rho(r) + \frac{d-2}{d-1} \frac{G Q^2}{2 \,\pi} \,\rho^2(r)+\mathcal{O}(G^2)\ ,
        \\
        \kappa\,h_{0 i}(\vec{x}) &= -2 (d-2) \frac{G m}{r^2}\rho(r)\,S_{ik}\,x^k + \frac{(d-2)^2}{d-1}\,\mathfrak{g}\,\frac{G Q^2}{4 \,\pi\, r^2}\rho^2(r)\,S_{ik}\,x^k+\mathcal{O}(G^2)\ ,
        \\
        \kappa\,h_{i j}(\vec{x}) &= -4 \frac{1}{d-1}G m\,\rho(r)\,\delta_{i j} + \frac{(d-3)\,r^2\,\delta_{i j}-(d-2)^2\,x_i x_j}{(d-1) (d-4)}\,\frac{G Q^2}{2 \, \pi \, r^2}\rho^2(r)+\mathcal{O}(G^2)\ .
    \end{aligned}
\end{equation}
Similarly, the electromagnetic potential is
\begin{equation}\label{eq:potential_d_dimensions}
    \begin{aligned}
        A_{0}(\vec{x}) &= \frac{Q}{4\,\pi}\,\rho(r)-\frac{d-2}{d-1}\frac{G m Q}{2 \,\pi}\,\rho^2(r)+\mathcal{O}(G^2)\ ,
        \\
        A_{i}(\vec{x}) &= (d-2)\,\mathfrak{g}\,\frac{Q}{8\,\pi \,r^2}\,\rho(r)\,S_{i k}\,x^k-\frac{(d-2)^2}{(d-1)^2} \left(d\,\frac{d-1}{d-2}-\mathfrak{g} \right) \frac{G m Q}{4 \, \pi \, r^2} \rho^2(r)\,S_{ik}\,x^k+\mathcal{O}(G^2)\ .
    \end{aligned}
\end{equation}
Equations \eqref{eq:metric_d_dimensions} and \eqref{eq:potential_d_dimensions}, therefore describe the metric and the electromagnetic potential of a charged spin-$\frac{1}{2}$ source, with an arbitrary value of its gyromagnetic factor $\mathfrak{g}$.

As we can see in equation \eqref{eq:metric_d_dimensions}, one encounters divergencies at one-loop in $4+1$ dimensions and, by the same mechanism, at two loops in $3+1$ dimensions~\cite{Mougiakakos:2020laz,DOnofrio:2022cvn}.
These divergences are cured by the addition of non-minimal higher-derivative terms built from the Riemann tensor and derivatives of the massive field~\cite{Mougiakakos:2020laz,Goldberger:2004jt}
\begin{equation} 
{}^{\mathrm{nm}}\mathcal{L}_{\psi^2 h}
    = C\,R~D_\mu\bar{\psi}~D^\mu\psi\ ,
\end{equation}
where $C$ is some dimensionful constant.
This coupling generates the counterterms needed to cancel the divergences that we would encounter in these particular dimensions, as detailed in~\cite{Mougiakakos:2020laz,DOnofrio:2022cvn}, resulting in logarithmic terms after the renormalization procedure. 

In addition, in odd space-time dimensions ($d$ even), the most general lagrangian could also include the corresponding Chern-Simons term~\cite{Witten:1988hf,Birmingham:1991ty}.
Particularly, in five dimensions the addition of this term is motivated by the arguments presented in~\cite{Chong:2005hr,Deshpande:2024vbn}.
Therefore, we write the five-dimensional Chern-Simons interaction as~\cite{Kunz:2017pnm,Deshpande:2024vbn}
\begin{equation}\label{eq:chern_simons_five_D}
    \mathcal{L}_{\mathrm{CS}_{d=4}}
    = \lambda\,\frac{\kappa}{16\sqrt{6}}\,\varepsilon^{\mu \nu \alpha \beta\gamma}\,F_{\mu \nu}\,F_{\alpha \beta}\,A_\gamma\ ,
\end{equation}
with $\lambda$ being a dimensionless free coefficient.
In the context of five-dimensional supergravity this Chern-Simons term \eqref{eq:chern_simons_five_D} is included, with $\lambda=1$, as the background theory to derive the solution presented in~\cite{Cvetic:2004hs,Chong:2005hr}.
In Refs.~\cite{Blazquez-Salcedo:2013wka,Kunz:2017pnm} they explore some properties of BHs with a different value for this parameter.

The insertion of the Chern-Simons term \eqref{eq:chern_simons_five_D} provides a new three-photon vertex
\begin{equation}\label{eq:chern_simons_vertex}
\includegraphics[valign=c,width=0.2\textwidth]{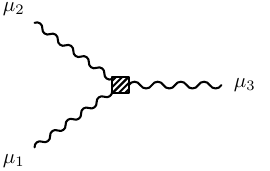} = \left(\tau_{A^3}\right)^{\mu_1 \mu_2 \mu_3} = -i\,\lambda\,\kappa\,\frac{\sqrt{6}}{4}\varepsilon^{\mu_1 \mu_2 \alpha \beta \mu_3}\,p_\alpha\,p'_\beta\ ,
\end{equation}
where $p_\alpha$ is an incoming momentum and $p'_{\beta}$ is an outgoing momentum.

\begin{figure}[h]
\centering
\begin{equation*}
    \includegraphics[valign=c,width=0.25\textwidth]{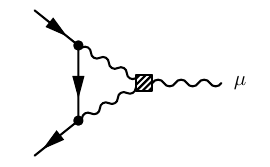}+2\times\includegraphics[valign=c,width=0.25\textwidth]{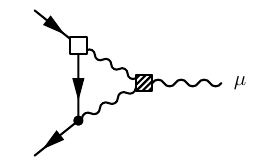}
\end{equation*}
\caption{Diagrams including the Chern-Simons interaction (the 3-photon squared vertex) with the corresponding multiplicity, inserted in five dimensions to compute up to one-loop and 1PM in dipole approximation.}
    \label{fig:potential_two}
\end{figure}

At one-loop the new Chern-Simons vertex is seen to only modify the electromagnetic potential.
Evaluating the diagrams seen in Fig. \ref{fig:potential_two} we get the following contribution
\begin{equation}\label{eq:02_potential_nm_and_nm2}
         A^{(\mathrm{CS})}_{\mu}(\vec{x}){\Bigg |}_{d=4} = - \lambda \,  \mathfrak{g} \,\frac{\sqrt{G}\,Q^2}{16\,\sqrt3\,\pi^{7/2}\,r^6}\,\varepsilon_{0 j k l \mu}\,S^{j k}\,x^l\ ,
\end{equation}
that must be added to Eq.~\eqref{eq:potential_d_dimensions} in order to get the full 1-loop expression for the potential.
In contrast, one expects the effects of \eqref{eq:chern_simons_five_D} to appear in the metric at two-loop order. 
The first contribution would be proportional to $Q^3 \, \sqrt{G^3}$, and hence be present at higher orders in the PM expansion.

\section{Particular cases}

In this section we compare our amplitude-derived metric and gauge potential with the PM expansion of known electrovacuum solutions in $3+1$ and $4+1$ dimensions.
We show that the Kerr-Newman solution in four dimensions corresponds to the minimally coupled EFT, whereas the five dimensional CCLP solution with null cosmological constant requires the non-minimal Pauli interaction.

\subsection{Kerr-Newman solution in $3+1$ dimensions}

 The Kerr-Newman solution describes the exterior of an asymptotically flat, stationary, charged and rotating BH in four dimensional Einstein-Maxwell theory~\cite{Newman:1965my,Adamo:2014baa}.
 In Boyer-Lindquist coordinates~\cite{Boyer:1966qh} $(t,\tilde{r},\theta,\phi)$, the metric reads~\cite{Frolov:1998wf}
\begin{equation}\label{eq:KN_metric_solution}
    \ddd s^2 = \frac{\Delta}{\Sigma}\left(\ddd t - a \sin^2\theta \, \ddd \varphi\right)^2 - \frac{\sin^2\theta}{\Sigma}\left(a\, \ddd t - (\tilde{r}^2+a^2)\,\ddd \varphi\right)^2-\frac{\Sigma}{\Delta} \ddd \tilde{r}^2-\Sigma\, \ddd\theta^2\ ,
\end{equation}
and the electromagnetic potential is
\begin{equation}\label{eq:KN_potential_solution}
    A_\mu\ddd x^\mu = \frac{Q}{4\pi}\frac{\tilde{r}}{\Sigma}\left( \ddd t -a \sin^2 \theta\,\ddd \varphi\right)\ ,
\end{equation}
where 
\begin{equation}
    \Sigma=\tilde{r}^2+a^2 \cos^2 \theta\ , \qquad \Delta = \tilde{r}^2+a^2-2 m G \tilde{r}+\frac{1}{4\pi} G Q^2
\end{equation}
and $a$ the standard spin-density parameter.

Transforming to Cartesian harmonic coordinates $(T,X,Y,Z)$ and expanding up to 1PM and dipole order we have
\begin{equation}\label{eq:kn_metric_pm_expansion}
    \begin{aligned}
        g^{\mathrm{KN}}_{00} &= 1 - 2 \frac{G m}{r} + \frac{G Q^2}{4\pi r^2}+ \mathcal{O}(G^2,Q^3,a^2)\ ,
	\\
	g^{\mathrm{KN}}_{0i} &= -2 \frac{G m}{r^3} \,S_{ik}\, x^k + \frac{G Q^2}{4 \pi r^4} \,S_{ik} \,x^k+ \mathcal{O}(G^2,Q^3,a^2)\ ,
	\\
	g^{\mathrm{KN}}_{ij} &= -\delta_{ij}-2 \frac{G m}{r}\,\delta_{ij}+\frac{G Q^2}{4 \pi r^4}\,x_i x_j + \mathcal{O}(G^2,Q^3,a^2)\ ,
    \end{aligned}
\end{equation}
and
\begin{equation}\label{eq:kn_potential_pm_expansion}
    \begin{aligned}
        A^{\mathrm{KN}}_{0} &= \frac{Q}{4 \, \pi \, r} - \frac{G m Q}{4 \, \pi \, r^2} + \mathcal{O}(G^2,Q^3,a^2)\ ,
	\\
        A^{\mathrm{KN}}_{i} &= \frac{Q}{4\, \pi \, r^3}\,S_{ik}\, x^k-\frac{G m Q}{4 \, \pi \, r^4}\,S_{ik}\,x^k+\mathcal{O}(G^2,Q^3,a^2)\ .
    \end{aligned}
\end{equation}

Then, setting $d=3$ in our general amplitude-based calculation, \eqref{eq:metric_d_dimensions} and \eqref{eq:potential_d_dimensions}, gives
\begin{equation}\label{eq:metric_3_dimensions}
    \begin{aligned}
        \left.\kappa\,h_{0 0}(\vec{x})\right|_{d=3} &= -2\frac{G m}{r} + \frac{G Q^2}{4 \, \pi \, r^2}+ \mathcal{O}(G^2,Q^3,a^2)\ ,
        \\
        \left.\kappa\,h_{0 i}(\vec{x})\right|_{d=3} &= -2 \frac{G m}{r^3}\,S_{ik}\,x^k + \mathfrak{g}\,\frac{G Q^2}{8 \, \pi \, r^4}\,S_{ik}\,x^k+ \mathcal{O}(G^2,Q^3,a^2)\ ,
        \\
        \left.\kappa\,h_{i j}(\vec{x})\right|_{d=3} &= -2 \frac{G m}{r}\,\delta_{i j} + \frac{G Q^2}{4 \, \pi \, r^4}\,x_i x_j+ \mathcal{O}(G^2,Q^3,a^2)\ ,
    \end{aligned}
\end{equation}
\begin{equation}\label{eq:potential_3_dimensions}
    \begin{aligned}
        \left.A_{0}(\vec{x})\right|_{d=3} &= \frac{Q}{4 \, \pi \, r}-\frac{G m Q}{4 \, \pi \, r^2}+ \mathcal{O}(G^2,Q^3,a^2)\ ,
        \\
        \left.A_{i}(\vec{x})\right|_{d=3} &= \mathfrak{g}\,\frac{Q}{8 \, \pi \, r^3} \,S_{ik}\,x^k- \left(6-\mathfrak{g} \right) \frac{G m Q}{16 \, \pi \, r^4} \,S_{ik}\,x^k+ \mathcal{O}(G^2,Q^3,a^2)\ .
    \end{aligned}
\end{equation}
Comparison with the Kerr-Newman far-field behavior shows exact agreement at 1PM and dipole order if and only if $\mathfrak{g}=2$ (equivalent to $\zeta=0$).
Thus, the Kerr-Newman metric is reproduced by the Dirac minimally coupled EFT of gravity at the order considered. This is in perfect agreement with the spinor-helicity analysis performed in~\cite{Moynihan:2019bor, Chung:2019yfs}.

\subsection{Chong-Cvetič-Lü-Pope solution in $4+1$ dimensions}

Charged rotating BHs in pure five-dimensional Einstein-Maxwell theory remain unknown in closed form~\cite{Kleihaus:2007kc}.
However, in five-dimensional gauged supergravity (which implies the inclusion of the specific Chern-Simons term given in \eqref{eq:chern_simons_five_D} with $\lambda=1$) one can construct the CCLP solution which is exact and will be given in the following~\cite{Cvetic:2004hs,Chong:2005hr}. Then, we will compare our amplitude-based result with the CCLP solution.

In Boyer-Lindquist type coordinates $(t,\tilde{r},\theta,\phi,\psi)$ and vanishing cosmological constant, the metric reads
\begin{equation}\label{eq:cclp_metric_solution}
    \begin{aligned}
        \ddd s^2 =& ~\ddd t^2+\frac{2\,\mathfrak{Q}}{\Sigma} \, \ddd\nu\, \ddd t-\frac{2 \mathfrak{Q}}{\Sigma}\, \ddd\nu\, \ddd\mu - \frac{\mathfrak{m}\, \tilde{r}^2-\mathfrak{Q}^2}{\Sigma^2}\,\left(\ddd t-\ddd\mu \right)^2-\frac{\Sigma}{\Delta} \ddd \tilde{r}^2
    \\
    &-\Sigma \,\ddd \theta^2-\left(\tilde{r}^2+a^2 \right) \sin^2\theta \,\ddd \phi^2-\left(\tilde{r}^2+b^2 \right) \cos^2\theta\, \ddd\psi^2\ ,
    \end{aligned}
\end{equation}
where one defines
\begin{equation}
    \begin{aligned}
    & \ddd\nu=b \sin^2\theta \,\ddd\phi+a \cos^2\theta \,\ddd\psi\ , \qquad
    \ddd\mu=a \sin^2\theta\,\ddd \phi + b \cos^2\theta \,\ddd\psi\ , \\
    \Sigma=\tilde{r}^2+a^2& \cos^2\theta + b^2 \sin^2 \theta\ , \qquad
    \Delta=\frac{1}{\tilde{r}^2}\left[(\tilde{r}^2+a^2)(\tilde{r}^2+b^2)+\mathfrak{Q}^2+2\, a\, b\, \mathfrak{Q} - \mathfrak{m}\, \tilde{r}^2\right]\ ,
    \end{aligned}
\end{equation}
and the gauge potential is
\begin{equation}\label{eq:cclp_potential_solution}
    A_\mu \ddd x^\mu = \sqrt{\frac{3}{\pi \,G}}\frac{\mathfrak{Q}}{4\, \Sigma} {\Big (} \ddd t - a \sin^2\theta \, \ddd \phi -b \cos^2\theta \,\ddd\psi{\Big )}\ .
\end{equation}
The parameters, $\mathfrak{m}$, $\mathfrak{Q}$, $a$ and $b$ are related to the physical mass, electric charge, and the two independent angular momenta by the relations~\cite{Chong:2005hr,Aliev:2006yk}
\begin{equation}\label{eq:CCLP_parameter_physical_relations}
    \mathfrak{m}=\frac{8 \,G \, m}{3\,\pi}\ , \quad \mathfrak{Q} = Q\, \sqrt{\frac{G}{3\, \pi^3}}\ , \quad J_1=\frac{\pi}{4\,G}{\Big (}a\, \mathfrak{m} + b\,\mathfrak{Q}{\Big )}\ , \quad J_2=\frac{\pi}{4 \,G} {\Big (}b\, \mathfrak{m} + a\, \mathfrak{Q}{\Big )}\ .
\end{equation}
It is worth noticing that the relations between the rotation parameters and the physical angular momenta in equation~\eqref{eq:CCLP_parameter_physical_relations} include an additional term, when compared to the standard Myers-Perry case~\cite{Myers:1986un}.
This extra contribution arises solely from the inclusion of the Chern-Simons interaction, which is proportional to the electric charge and induces a cross-relation between the rotation parameters $a,b$ and the physical angular momenta $J_1, J_2$.

To proceed as in the Kerr-Newman case, we first move to the Cartesian harmonic coordinates. As usual, we choose  the plane $(X_i,Y_i)$ to be orthogonal to the angular momentum $J_i$, and this  results in a block diagonal spin-density tensor
\begin{equation}
    S_{i j} = \left(
    \begin{array}{cccc}
         0&J_1/m&0&0  
         \\
         -J_1/m&0&0&0 
         \\
         0&0&0&J_2/m
         \\
         0&0&-J_2/m&0
    \end{array}
    \right)\ .
\end{equation}
Then, expanding up to 1PM and dipole order the CCLP solution, we find 
\begin{equation}\label{eq:CCLP_solution_pm_expansion_metric}
    \begin{aligned}
    g^{\mathrm{CCLP}}_{00} =&~ 1 - \frac{8\, G m}{3 \, \pi \, r^2} + \frac{G Q^2}{3 \, \pi^3 \, r^4}+ \mathcal{O}(G^2,Q^3,S^2)\ ,
	\\
	g^{\mathrm{CCLP}}_{0i} =& - \frac{4\,G m}{\pi \, r^3} \,S_{ik}\,x^k + \frac{G Q^2}{2 \, \pi^3 \, r^6} \,S_{ik}\,x^k+ \mathcal{O}(G^2,Q^3,S^2)\ ,
	\\
	g^{\mathrm{CCLP}}_{ij} =& -\delta_{ij}- \frac{4\,G m}{3 \, \pi \, r^2}\,\delta_{ij}+\frac{G Q^2}{12 \, \pi^3 \, r^4}\,\log\left( \frac{8\,G m}{3 \, \pi \, r^2} \right)\,\delta_{ij}
    \\
    &+\frac{G Q^2}{6 \, \pi^3 \, r^6}\,x_i x_j-\frac{G Q^2}{3 \, \pi^3 \, r^6}\,\log\left( \frac{8\,G m}{3 \, \pi \, r^2} \right)\,x_i x_j + \mathcal{O}(G^2,Q^3,S^2)\ ,
    \end{aligned}
\end{equation}
and
\begin{equation}\label{eq:CCLP_solution_pm_expansion_potential}
    \begin{aligned}
    A^{\mathrm{CCLP}}_{0} =&~ \frac{Q}{4\, \pi^2 \, r^2} - \frac{G m Q}{3 \, \pi^3 \, r^4} + \mathcal{O}(G^2,Q^3,S^2) ,
    \\
    A^{\mathrm{CCLP}}_{i} =&~ \frac{3\,Q}{8 \, \pi^2 \, r^4}\,S_{ik}\,x^k-\frac{G m Q}{2 \, \pi^3 \, r^6}\,S_{ik}\,x^k
    \\
    &- \frac{\sqrt{3}\,\sqrt{G}\,Q^2}{32\, \pi^{7/2} \,r^6}\,\varepsilon_{0 j k l i}\,S^{j k}\,x^l+ \mathcal{O}(G^2,Q^3,S^2)\ .
    \end{aligned}
\end{equation}

On the other hand, adding the contribution from \eqref{eq:02_potential_nm_and_nm2} and performing the renormalization procedure~\cite{DOnofrio:2022cvn} for $d=4$, our amplitude-based calculation in Eqs.~\eqref{eq:metric_d_dimensions} and \eqref{eq:potential_d_dimensions} becomes
\begin{equation}\label{eq:metric_5_dimensions}
    \begin{aligned}
        \kappa\,h_{0 0}(\vec{x})\Big|_{d=4} &= - \frac{8\, G m}{3 \, \pi \, r^2} + \frac{G Q^2}{3 \, \pi^3 \, r^4}+ \mathcal{O}(G^2,Q^3,S^2)\ , \\
        \kappa\,h_{0 i}(\vec{x})\Big|_{d=4} &= - \frac{4\,G m}{\pi\, r^3} \,S_{ik}\,x^k + \mathfrak{g}\,\frac{G Q^2}{3\, \pi^3 \,r^6}\,S_{ik}\,x^k+ \mathcal{O}(G^2,Q^3,S^2)\ ,\\
        \kappa\,h_{i j}(\vec{x})\Big|_{d=4} &= - \frac{4\,G m}{3\,\pi \,r^2}\,\delta_{ij}+\frac{G Q^2}{12\, \pi^3 \,r^4}\,\log\left( \frac{8\,G m}{3 \, \pi\, r^2} \right)\,\delta_{ij}\\
        &+\frac{G Q^2}{6 \, \pi^3 \, r^6}\,x_i x_j-\frac{G Q^2}{3 \,\pi^3\, r^6}\,\log\left( \frac{8\,G m}{3 \,\pi \,r^2} \right)\,x_i x_j+ \mathcal{O}(G^2,Q^3,S^2)\ ,
    \end{aligned}
\end{equation}
\begin{equation}\label{eq:potential_5_dimensions}
    \begin{aligned}
        A_{0}(\vec{x})\Big|_{d=4} &= \frac{Q}{4\, \pi^2 \,r^2} - \frac{G m Q}{3\, \pi^3 \,r^4}+ \mathcal{O}(G^2,Q^3,S^2)\ ,\\
        A_{i}(\vec{x})\Big|_{d=4} &= \mathfrak{g}\,\frac{Q}{4\,\pi^2 \,r^4} \,S_{ik}\,x^k- \left(6-\mathfrak{g} \right) \frac{G m Q}{9\, \pi^3 \,r^6}\,S_{ik}\,x^k\\
        &- \lambda \,  \mathfrak{g} \,\frac{\sqrt{G}\,Q^2}{16\,\sqrt3\,\pi^{7/2}\,r^6}\,\varepsilon_{0 j k l i}\,S^{j k}\,x^l+ \mathcal{O}(G^2,Q^3,S^2)\ .
    \end{aligned}
\end{equation}

Comparing the amplitude-derived expression with the expansion of the CCLP solution fixes the gyromagnetic factor to be $\mathfrak{g}=\frac{3}{2}$, corresponding to the inclusion of a Pauli coupling with parameter $\zeta=-\frac{1}{4}$.
As expected, the value $\lambda=1$ is needed to match the two descriptions, since the solution originates from five-dimensional supergravity.

\subsection{Gyromagnetic factor of higher-dimensional black holes}

To analyze the dimensional dependence of the gyromagnetic factor of BHs, we consider the work approached in~\cite{Aliev:2004ec,Aliev:2006yk}, where they use the Myers-Perry solution as a chargeless rotating background in higher dimensions. Then, a small electric charge is introduced, allowing to perturbatively construct an electromagnetic potential, by solving the Einstein-Maxwell equations, associated with the slow-rotating and small-charged BH in general dimensions.

In the same way, since we are only interested in the dipole order term of the solution, we take the following ansatz for the magnetic part in dipole approximation of the gauge potential in Cartesian-like coordinates
\begin{equation}\label{eq:potential_ansatz_q}
    A_i=\frac{q}{r^d}\,\frac{d-1}{2}\,S_{ik}\,x^k\ ,
\end{equation}
where we have used the usual relation between the rotation parameters and the physical angular momenta in higher dimensions $a_i=\frac{d-1}{2}\frac{J_i}{m}$~\cite{Myers:1986un}, and $q$ is a parameter related to the physical electric charge $Q$, defined via the Komar integral at infinity as
\begin{equation}
    Q =  \int_{\Omega_{d-1}} \star\, F =   \, q\,(d-2)\,\Omega_{d-1}\ ,
\end{equation}
where $\star\, F$ is the Hodge dual of the electromagnetic field strength tensor.
Solving for $q$ and substituting it back into \eqref{eq:potential_ansatz_q}, we obtain the magnetic part of the charged BH in terms of the physical charge
\begin{equation}\label{eq:bh_d_g}
    A_i=\frac{1}{\Omega_{d-1}~r^d}\,\frac{d-1}{d-2}\, \frac{Q}{2}S_{ik}\,x^k\ .
\end{equation}
The form of the potential shown in \eqref{eq:bh_d_g} recovers \eqref{eq:kn_potential_pm_expansion} setting $d=3$, and also \eqref{eq:CCLP_solution_pm_expansion_potential} taking $d= 4$.

Comparing \eqref{eq:bh_d_g} with our amplitude-based result \eqref{eq:dipole_potential}, we find that the gyromagnetic factor of a charged Myers-Perry like BH in $d+1$ dimensions is
\begin{equation}\label{eq:general_d_bh_g}
    \mathfrak{g}_{\mathrm{BH}} = \frac{d-1}{d-2}\ ,
\end{equation}
therefore reproducing the $\mathfrak{g}_{\mathrm{BH}}{\big |}_{d=3} = 2$ and $\mathfrak{g}_{\mathrm{BH}}{\big |}_{d=4} = \frac{3}{2}$ values that we previously deduced from the Kerr-Newman and CCLP examples. Notice that Eq.~\eqref{eq:general_d_bh_g} coincides with the expression derived in~\cite{Ortaggio:2006ng} up to a normalization factor. 

Hence, only in four dimensions the BH PM expansion can be recovered from a minimally coupled EFT of gravity, while for $d>3$ one must include a Pauli-type coupling with a dimensionless coefficient $\zeta = -\frac{d-3}{2\,(d-2)}$.

\section{Discussion}

In this paper we have shown how to reconstruct the far-field metric and electromagnetic potential of charged, rotating BHs in higher dimensions. We have done it by matching graviton and photon emission scattering amplitudes of a massive charged spin-$\frac{1}{2}$ source to the PM expansions of known BH solutions. 
We carried out the calculation up to 1-loop order and in dipole approximation, thereby extending the works~\cite{Donoghue:2001qc} and~\cite{DOnofrio:2022cvn}, namely generalizing them to higher dimensions and including rotating sources.

Within our approach and approximations we reproduced the Kerr-Newman and the CCLP solutions. In the former case we have shown that the such solution emerges from a minimally-coupled EFT, while in the latter case we need to introduce a non-minimally Pauli-like interaction term and a specific Chern-Simons coupling. Such coupling is precisely the one appearing in the five-dimensional gauged supergravity of which the CCLP BH is a solution.   

For $\lambda\neq 1$ an exact solution no longer exists, however in~\cite{Blazquez-Salcedo:2013wka,Kleihaus:2007kc,Kunz:2017pnm, Ortaggio:2023rzp} they analyze the effects of this parameter within a near-horizon expansion. Comparing these studies with our far-field analysis could provide new insights on how the dimensionless Chern-Simons coefficient $\lambda$ enters into the structure of the solution changing the relation between the physical angular momentum with respect to the rotation parameters.

Finally, the main insight of this work is that, by introducing a Pauli-like interaction we probed the magnetic properties of these higher-dimensional BHs.
Matching the amplitude-based computations to far-field asymptotic behavior of charged Myers-Perry like solutions in $d+1$ dimensions gives a gyromagnetic factor of $\mathfrak{g}_{\mathrm{BH}}=\frac{d-1}{d-2}$, which in EFT language corresponds to the introduction of a Pauli parameter $\zeta=-\frac{d-3}{2\,(d-2)}$. Notably, this result is independent of the presence of a Chern-Simons term. This universality arises because such interaction does not contribute to the electromagnetic dipole moment of the solution.
Consequently, our general result applies equally to the rotating BH solutions of~\cite{Aliev:2004ec,Aliev:2006yk} and those of~\cite{Chong:2005hr,Kunz:2017pnm}.

The natural next step is to push our framework to quadrupole order, for instance, by employing charged spin-$1$ fields, where even gravitational multipoles became non-trivial, as already explored in the chargeless higher-dimensional case~\cite{Kleihaus:2007kc,Gambino:2024uge}. Moreover it would be interesting to compare our analysis with the results in~\cite{Deshpande:2024vbn} where the CCLP solution is extended to odd spacetime dimensions higher than $D=5$.

\begin{acknowledgments}
We would like to thank P. Pani for discussions. 
This work is partially supported by Sapienza University of Rome (``Progetti per Avvio alla Ricerca - Tipo 1'',
protocol number AR1241906DC8FF32).
\end{acknowledgments}

\bibliographystyle{JHEP}
\bibliography{biblio}

\providecommand{\href}[2]{#2}\begingroup\raggedright\begin{thebibliography}{10}

\bibitem{Bardeen:1973gs}
J.M.~Bardeen, B.~Carter and S.W.~Hawking, \emph{{The Four laws of black hole
  mechanics}}, \href{https://doi.org/10.1007/BF01645742}{\emph{Commun. Math.
  Phys.} {\bfseries 31} (1973) 161}.

\bibitem{Robinson:1975bv}
D.C.~Robinson, \emph{{Uniqueness of the Kerr black hole}},
  \href{https://doi.org/10.1103/PhysRevLett.34.905}{\emph{Phys. Rev. Lett.}
  {\bfseries 34} (1975) 905}.

\bibitem{Robinson:2004zz}
D.~Robinson, \emph{{Four decades of black holes uniqueness theorems}},  in
  \emph{{Kerr Fest: Black Holes in Astrophysics, General Relativity and Quantum
  Gravity}}, 8, 2004.

\bibitem{Cardoso:2016ryw}
V.~Cardoso and L.~Gualtieri, \emph{{Testing the black hole
  {\textquoteleft}no-hair{\textquoteright} hypothesis}},
  \href{https://doi.org/10.1088/0264-9381/33/17/174001}{\emph{Class. Quant.
  Grav.} {\bfseries 33} (2016) 174001}
  [\href{https://arxiv.org/abs/1607.03133}{{\ttfamily 1607.03133}}].

\bibitem{Geroch:1970cd}
R.P.~Geroch, \emph{{Multipole moments. II. Curved space}},
  \href{https://doi.org/10.1063/1.1665427}{\emph{J. Math. Phys.} {\bfseries 11}
  (1970) 2580}.

\bibitem{Thorne:1980ru}
K.S.~Thorne, \emph{{Multipole Expansions of Gravitational Radiation}},
  \href{https://doi.org/10.1103/RevModPhys.52.299}{\emph{Rev. Mod. Phys.}
  {\bfseries 52} (1980) 299}.

\bibitem{Hansen:1974zz}
R.O.~Hansen, \emph{{Multipole moments of stationary space-times}},
  \href{https://doi.org/10.1063/1.1666501}{\emph{J. Math. Phys.} {\bfseries 15}
  (1974) 46}.

\bibitem{Bern:2019nnu}
Z.~Bern, C.~Cheung, R.~Roiban, C.-H.~Shen, M.P.~Solon and M.~Zeng,
  \emph{{Scattering Amplitudes and the Conservative Hamiltonian for Binary
  Systems at Third Post-Minkowskian Order}},
  \href{https://doi.org/10.1103/PhysRevLett.122.201603}{\emph{Phys. Rev. Lett.}
  {\bfseries 122} (2019) 201603}
  [\href{https://arxiv.org/abs/1901.04424}{{\ttfamily 1901.04424}}].

\bibitem{Damour:2016gwp}
T.~Damour, \emph{{Gravitational scattering, post-Minkowskian approximation and
  Effective One-Body theory}},
  \href{https://doi.org/10.1103/PhysRevD.94.104015}{\emph{Phys. Rev. D}
  {\bfseries 94} (2016) 104015}
  [\href{https://arxiv.org/abs/1609.00354}{{\ttfamily 1609.00354}}].

\bibitem{Gambino:2024uge}
C.~Gambino, P.~Pani and F.~Riccioni, \emph{{Rotating metrics and new multipole
  moments from scattering amplitudes in arbitrary dimensions}},
  \href{https://doi.org/10.1103/PhysRevD.109.124018}{\emph{Phys. Rev. D}
  {\bfseries 109} (2024) 124018}
  [\href{https://arxiv.org/abs/2403.16574}{{\ttfamily 2403.16574}}].

\bibitem{Misner:1973prb}
C.W.~Misner, K.S.~Thorne and J.A.~Wheeler, \emph{{Gravitation}}, W. H. Freeman,
  San Francisco (1973).

\bibitem{Kleihaus:2007kc}
B.~Kleihaus, J.~Kunz and F.~Navarro-Lerida, \emph{{Rotating Black Holes in
  Higher Dimensions}}, \href{https://doi.org/10.1063/1.2902801}{\emph{AIP Conf.
  Proc.} {\bfseries 977} (2008) 94}
  [\href{https://arxiv.org/abs/0710.2291}{{\ttfamily 0710.2291}}].

\bibitem{Schwarzschild:1916uq}
K.~Schwarzschild, \emph{{On the gravitational field of a mass point according
  to Einstein's theory}}, {\emph{Sitzungsber. Preuss. Akad. Wiss. Berlin (Math.
  Phys. )} {\bfseries 1916} (1916) 189}
  [\href{https://arxiv.org/abs/physics/9905030}{{\ttfamily physics/9905030}}].

\bibitem{Reissner:1916cle}
H.~Reissner, \emph{{{\"U}ber die Eigengravitation des elektrischen Feldes nach
  der Einsteinschen Theorie}},
  \href{https://doi.org/10.1002/andp.19163550905}{\emph{Annalen Phys.}
  {\bfseries 355} (1916) 106}.

\bibitem{Kerr:1963ud}
R.P.~Kerr, \emph{{Gravitational field of a spinning mass as an example of
  algebraically special metrics}},
  \href{https://doi.org/10.1103/PhysRevLett.11.237}{\emph{Phys. Rev. Lett.}
  {\bfseries 11} (1963) 237}.

\bibitem{Newman:1965my}
E.T.~Newman, E.~Couch, K.~Chinnapared, A.~Exton, A.~Prakash and R.~Torrence,
  \emph{{Metric of a Rotating, Charged Mass}},
  \href{https://doi.org/10.1063/1.1704351}{\emph{J. Math. Phys.} {\bfseries 6}
  (1965) 918}.

\bibitem{Emparan:2008eg}
R.~Emparan and H.S.~Reall, \emph{{Black Holes in Higher Dimensions}},
  \href{https://doi.org/10.12942/lrr-2008-6}{\emph{Living Rev. Rel.} {\bfseries
  11} (2008) 6} [\href{https://arxiv.org/abs/0801.3471}{{\ttfamily
  0801.3471}}].

\bibitem{Myers:1986un}
R.C.~Myers and M.J.~Perry, \emph{{Black Holes in Higher Dimensional
  Space-Times}},
  \href{https://doi.org/10.1016/0003-4916(86)90186-7}{\emph{Annals Phys.}
  {\bfseries 172} (1986) 304}.

\bibitem{Emparan:2001wn}
R.~Emparan and H.S.~Reall, \emph{{A Rotating black ring solution in
  five-dimensions}},
  \href{https://doi.org/10.1103/PhysRevLett.88.101101}{\emph{Phys. Rev. Lett.}
  {\bfseries 88} (2002) 101101}
  [\href{https://arxiv.org/abs/hep-th/0110260}{{\ttfamily hep-th/0110260}}].

\bibitem{Tangherlini:1963bw}
F.R.~Tangherlini, \emph{{Schwarzschild field in n dimensions and the
  dimensionality of space problem}},
  \href{https://doi.org/10.1007/BF02784569}{\emph{Nuovo Cim.} {\bfseries 27}
  (1963) 636}.

\bibitem{Kunz:2017pnm}
J.~Kunz, J.L.~Bl{\'a}zquez-Salcedo, F.~Navarro-L{\'e}rida and E.~Radu,
  \emph{{Einstein-Maxwell-Chern-Simons Black Holes}},
  \href{https://doi.org/10.1088/1742-6596/942/1/012003}{\emph{J. Phys. Conf.
  Ser.} {\bfseries 942} (2017) 012003}
  [\href{https://arxiv.org/abs/1709.09552}{{\ttfamily 1709.09552}}].

\bibitem{Ortaggio:2023rzp}
M.~Ortaggio and A.~Srinivasan, \emph{{Charging Kerr-Schild spacetimes in higher
  dimensions}}, \href{https://doi.org/10.1103/PhysRevD.110.044035}{\emph{Phys.
  Rev. D} {\bfseries 110} (2024) 044035}
  [\href{https://arxiv.org/abs/2309.02900}{{\ttfamily 2309.02900}}].

\bibitem{Aliev:2004ec}
A.N.~Aliev and V.P.~Frolov, \emph{{Five-dimensional rotating black hole in a
  uniform magnetic field: The Gyromagnetic ratio}},
  \href{https://doi.org/10.1103/PhysRevD.69.084022}{\emph{Phys. Rev. D}
  {\bfseries 69} (2004) 084022}
  [\href{https://arxiv.org/abs/hep-th/0401095}{{\ttfamily hep-th/0401095}}].

\bibitem{Navarro-Lerida:2010orf}
F.~Navarro-Lerida, \emph{{Perturbative Charged Rotating 5D Einstein-Maxwell
  Black Holes}}, \href{https://doi.org/10.1007/s10714-010-1033-1}{\emph{Gen.
  Rel. Grav.} {\bfseries 42} (2010) 2891}
  [\href{https://arxiv.org/abs/0706.0591}{{\ttfamily 0706.0591}}].

\bibitem{Aliev:2006yk}
A.N.~Aliev, \emph{{Rotating black holes in higher dimensional Einstein-Maxwell
  gravity}}, \href{https://doi.org/10.1103/PhysRevD.74.024011}{\emph{Phys. Rev.
  D} {\bfseries 74} (2006) 024011}
  [\href{https://arxiv.org/abs/hep-th/0604207}{{\ttfamily hep-th/0604207}}].

\bibitem{Horowitz:1995tm}
G.T.~Horowitz and A.~Sen, \emph{{Rotating black holes which saturate a
  Bogomolny bound}}, \href{https://doi.org/10.1103/PhysRevD.53.808}{\emph{Phys.
  Rev. D} {\bfseries 53} (1996) 808}
  [\href{https://arxiv.org/abs/hep-th/9509108}{{\ttfamily hep-th/9509108}}].

\bibitem{Llatas:1996gh}
P.M.~Llatas, \emph{{Electrically charged black holes for the heterotic string
  compactified on a (10 - D) torus}},
  \href{https://doi.org/10.1016/S0370-2693(97)00144-5}{\emph{Phys. Lett. B}
  {\bfseries 397} (1997) 63}
  [\href{https://arxiv.org/abs/hep-th/9605058}{{\ttfamily hep-th/9605058}}].

\bibitem{Youm:1997hw}
D.~Youm, \emph{{Black holes and solitons in string theory}},
  \href{https://doi.org/10.1016/S0370-1573(99)00037-X}{\emph{Phys. Rept.}
  {\bfseries 316} (1999) 1}
  [\href{https://arxiv.org/abs/hep-th/9710046}{{\ttfamily hep-th/9710046}}].

\bibitem{Chong:2005hr}
Z.W.~Chong, M.~Cvetic, H.~Lu and C.N.~Pope, \emph{{General non-extremal
  rotating black holes in minimal five-dimensional gauged supergravity}},
  \href{https://doi.org/10.1103/PhysRevLett.95.161301}{\emph{Phys. Rev. Lett.}
  {\bfseries 95} (2005) 161301}
  [\href{https://arxiv.org/abs/hep-th/0506029}{{\ttfamily hep-th/0506029}}].

\bibitem{Kunz:2006jd}
J.~Kunz, D.~Maison, F.~Navarro-Lerida and J.~Viebahn, \emph{{Rotating
  Einstein-Maxwell-dilaton black holes in D dimensions}},
  \href{https://doi.org/10.1016/j.physletb.2006.06.024}{\emph{Phys. Lett. B}
  {\bfseries 639} (2006) 95}
  [\href{https://arxiv.org/abs/hep-th/0606005}{{\ttfamily hep-th/0606005}}].

\bibitem{Blazquez-Salcedo:2013wka}
J.L.~Blazquez-Salcedo, J.~Kunz and F.~Navarro-Lerida, \emph{{Properties of
  rotating Einstein-Maxwell-Dilaton black holes in odd dimensions}},
  \href{https://doi.org/10.1103/PhysRevD.89.024038}{\emph{Phys. Rev. D}
  {\bfseries 89} (2014) 024038}
  [\href{https://arxiv.org/abs/1311.0062}{{\ttfamily 1311.0062}}].

\bibitem{Deshpande:2024vbn}
R.~Deshpande and O.~Lunin, \emph{{Rotating Einstein-Maxwell black holes in odd
  dimensions}}, \href{https://doi.org/10.1007/JHEP06(2025)066}{\emph{JHEP}
  {\bfseries 06} (2025) 066}
  [\href{https://arxiv.org/abs/2411.01795}{{\ttfamily 2411.01795}}].

\bibitem{Chung:2019yfs}
M.-Z.~Chung, Y.-T.~Huang and J.-W.~Kim, \emph{{Kerr-Newman stress-tensor from
  minimal coupling}},
  \href{https://doi.org/10.1007/JHEP12(2020)103}{\emph{JHEP} {\bfseries 12}
  (2020) 103} [\href{https://arxiv.org/abs/1911.12775}{{\ttfamily
  1911.12775}}].

\bibitem{Moynihan:2019bor}
N.~Moynihan, \emph{{Kerr-Newman from Minimal Coupling}},
  \href{https://doi.org/10.1007/JHEP01(2020)014}{\emph{JHEP} {\bfseries 01}
  (2020) 014} [\href{https://arxiv.org/abs/1909.05217}{{\ttfamily
  1909.05217}}].

\bibitem{Chung:2018kqs}
M.-Z.~Chung, Y.-T.~Huang, J.-W.~Kim and S.~Lee, \emph{{The simplest massive
  S-matrix: from minimal coupling to Black Holes}},
  \href{https://doi.org/10.1007/JHEP04(2019)156}{\emph{JHEP} {\bfseries 04}
  (2019) 156} [\href{https://arxiv.org/abs/1812.08752}{{\ttfamily
  1812.08752}}].

\bibitem{Mougiakakos:2020laz}
S.~Mougiakakos and P.~Vanhove, \emph{{Schwarzschild-Tangherlini metric from
  scattering amplitudes in various dimensions}},
  \href{https://doi.org/10.1103/PhysRevD.103.026001}{\emph{Phys. Rev. D}
  {\bfseries 103} (2021) 026001}
  [\href{https://arxiv.org/abs/2010.08882}{{\ttfamily 2010.08882}}].

\bibitem{DOnofrio:2022cvn}
S.~D'Onofrio, F.~Fragomeno, C.~Gambino and F.~Riccioni, \emph{{The
  Reissner-Nordstr{\"o}m-Tangherlini solution from scattering amplitudes of
  charged scalars}}, \href{https://doi.org/10.1007/JHEP09(2022)013}{\emph{JHEP}
  {\bfseries 09} (2022) 013}
  [\href{https://arxiv.org/abs/2207.05841}{{\ttfamily 2207.05841}}].

\bibitem{Donoghue:1994dn}
J.F.~Donoghue, \emph{{General relativity as an effective field theory: The
  leading quantum corrections}},
  \href{https://doi.org/10.1103/PhysRevD.50.3874}{\emph{Phys. Rev. D}
  {\bfseries 50} (1994) 3874}
  [\href{https://arxiv.org/abs/gr-qc/9405057}{{\ttfamily gr-qc/9405057}}].

\bibitem{Bjerrum-Bohr:2002fji}
N.E.J.~Bjerrum-Bohr, J.F.~Donoghue and B.R.~Holstein, \emph{{Quantum
  corrections to the Schwarzschild and Kerr metrics}},
  \href{https://doi.org/10.1103/PhysRevD.68.084005}{\emph{Phys. Rev. D}
  {\bfseries 68} (2003) 084005}
  [\href{https://arxiv.org/abs/hep-th/0211071}{{\ttfamily hep-th/0211071}}].

\bibitem{Holstein:2004dn}
B.R.~Holstein and J.F.~Donoghue, \emph{{Classical physics and quantum loops}},
  \href{https://doi.org/10.1103/PhysRevLett.93.201602}{\emph{Phys. Rev. Lett.}
  {\bfseries 93} (2004) 201602}
  [\href{https://arxiv.org/abs/hep-th/0405239}{{\ttfamily hep-th/0405239}}].

\bibitem{Pauli:1941zz}
W.~Pauli, \emph{{Relativistic Field Theories of Elementary Particles}},
  \href{https://doi.org/10.1103/RevModPhys.13.203}{\emph{Rev. Mod. Phys.}
  {\bfseries 13} (1941) 203}.

\bibitem{Itzykson:1980rh}
C.~Itzykson and J.B.~Zuber, \emph{{Quantum Field Theory}}, International Series
  In Pure and Applied Physics, McGraw-Hill, New York (1980).

\bibitem{Kosower:2018adc}
D.A.~Kosower, B.~Maybee and D.~O'Connell, \emph{{Amplitudes, Observables, and
  Classical Scattering}},
  \href{https://doi.org/10.1007/JHEP02(2019)137}{\emph{JHEP} {\bfseries 02}
  (2019) 137} [\href{https://arxiv.org/abs/1811.10950}{{\ttfamily
  1811.10950}}].

\bibitem{Bern:2020buy}
Z.~Bern, A.~Luna, R.~Roiban, C.-H.~Shen and M.~Zeng, \emph{{Spinning black hole
  binary dynamics, scattering amplitudes, and effective field theory}},
  \href{https://doi.org/10.1103/PhysRevD.104.065014}{\emph{Phys. Rev. D}
  {\bfseries 104} (2021) 065014}
  [\href{https://arxiv.org/abs/2005.03071}{{\ttfamily 2005.03071}}].

\bibitem{Donoghue:2001qc}
J.F.~Donoghue, B.R.~Holstein, B.~Garbrecht and T.~Konstandin, \emph{{Quantum
  corrections to the Reissner-Nordstr{\"o}m and Kerr-Newman metrics}},
  \href{https://doi.org/10.1016/S0370-2693(02)01246-7}{\emph{Phys. Lett. B}
  {\bfseries 529} (2002) 132}
  [\href{https://arxiv.org/abs/hep-th/0112237}{{\ttfamily hep-th/0112237}}].

\bibitem{Bjerrum-Bohr:2018xdl}
N.E.J.~Bjerrum-Bohr, P.H.~Damgaard, G.~Festuccia, L.~Plant{\'e} and P.~Vanhove,
  \emph{{General Relativity from Scattering Amplitudes}},
  \href{https://doi.org/10.1103/PhysRevLett.121.171601}{\emph{Phys. Rev. Lett.}
  {\bfseries 121} (2018) 171601}
  [\href{https://arxiv.org/abs/1806.04920}{{\ttfamily 1806.04920}}].

\bibitem{Goldberger:2004jt}
W.D.~Goldberger and I.Z.~Rothstein, \emph{{An Effective field theory of gravity
  for extended objects}},
  \href{https://doi.org/10.1103/PhysRevD.73.104029}{\emph{Phys. Rev. D}
  {\bfseries 73} (2006) 104029}
  [\href{https://arxiv.org/abs/hep-th/0409156}{{\ttfamily hep-th/0409156}}].

\bibitem{Witten:1988hf}
E.~Witten, \emph{{Quantum Field Theory and the Jones Polynomial}},
  \href{https://doi.org/10.1007/BF01217730}{\emph{Commun. Math. Phys.}
  {\bfseries 121} (1989) 351}.

\bibitem{Birmingham:1991ty}
D.~Birmingham, M.~Blau, M.~Rakowski and G.~Thompson, \emph{{Topological field
  theory}}, \href{https://doi.org/10.1016/0370-1573(91)90117-5}{\emph{Phys.
  Rept.} {\bfseries 209} (1991) 129}.

\bibitem{Cvetic:2004hs}
M.~Cvetic, H.~Lu and C.N.~Pope, \emph{{Charged Kerr-de Sitter black holes in
  five dimensions}},
  \href{https://doi.org/10.1016/j.physletb.2004.08.011}{\emph{Phys. Lett. B}
  {\bfseries 598} (2004) 273}
  [\href{https://arxiv.org/abs/hep-th/0406196}{{\ttfamily hep-th/0406196}}].

\bibitem{Adamo:2014baa}
T.~Adamo and E.T.~Newman, \emph{{The Kerr-Newman metric: A Review}},
  \href{https://doi.org/10.4249/scholarpedia.31791}{\emph{Scholarpedia}
  {\bfseries 9} (2014) 31791}
  [\href{https://arxiv.org/abs/1410.6626}{{\ttfamily 1410.6626}}].

\bibitem{Boyer:1966qh}
R.H.~Boyer and R.W.~Lindquist, \emph{{Maximal analytic extension of the Kerr
  metric}}, \href{https://doi.org/10.1063/1.1705193}{\emph{J. Math. Phys.}
  {\bfseries 8} (1967) 265}.

\bibitem{Frolov:1998wf}
V.P.~Frolov and I.D.~Novikov, eds., \emph{{Black hole physics: Basic concepts
  and new developments}} (1998),
  \href{https://doi.org/10.1007/978-94-011-5139-9}{10.1007/978-94-011-5139-9}.

\bibitem{Ortaggio:2006ng}
M.~Ortaggio and V.~Pravda, \emph{{Black rings with a small electric charge:
  Gyromagnetic ratios and algebraic alignment}},
  \href{https://doi.org/10.1088/1126-6708/2006/12/054}{\emph{JHEP} {\bfseries
  12} (2006) 054} [\href{https://arxiv.org/abs/gr-qc/0609049}{{\ttfamily
  gr-qc/0609049}}].

\end{thebibliography}\endgroup
\end{document}